\documentclass[journal]{IEEEtran}
\usepackage{cite}
\usepackage{amsmath,amssymb,amsfonts}
\usepackage{algorithmic}
\usepackage{graphicx}
\usepackage{textcomp}
\usepackage{booktabs}
\usepackage{pifont}
\usepackage{crimson}
\usepackage{xspace}
\usepackage{caption,setspace}
\newcommand{\etal}{\emph{et al.}\xspace}
\newcommand{\mat}[1]{\boldsymbol{#1}} 
\def\BibTeX{{\rm B\kern-.05em{\sc i\kern-.025em b}\kern-.08em
    T\kern-.1667em\lower.7ex\hbox{E}\kern-.125emX}}
\usepackage{atbegshi}
\AtBeginDocument{\AtBeginShipoutNext{\AtBeginShipoutDiscard}}

\usepackage{url}
\begin{document}

\title{Deep Efficient End-to-end Reconstruction (DEER) Network for Few-view Breast CT
Image Reconstruction}
\author{Huidong Xie$^{1,2}$, Hongming Shan$^{1}$, Wenxiang Cong$^{1}$, Chi Liu$^{2,3}$, Xiaohua Zhang$^{4}$, Shaohua Liu$^{4}$, Ruola Ning$^{4}$, and Ge Wang$^{1}$}%

\thanks{$^1$Department of Biomedical Engineering, Biomedical Imaging Center, Center for Biotechnology and Interdisciplinary Studies, Rensselaer Polytechnic Institute, Troy, NY 12180, USA}%
\thanks{$^2$Department of Biomedical Engineering, Yale University, New Haven, CT 06520, USA}%
\thanks{$^3$Department of Radiology and Biomedical Imaging, Yale University, New Haven, CT 06520, USA}%
\thanks{$^4$Koning Corporation, West Henrietta, NY 14586, USA}%
\thanks{This work was supported in part by the NIH/NCI under Award R01CA233888 and Award R01CA237267, in part by the NIH/NIBIB under Award R01EB026646, and in part by the NIH/NHLBI under Award R01HL151561.}%
\thanks{Corresponding authors: Hongming Shan (e-mail: hmshan@ieee.org), and Ge Wang (e-mail: wangg6@rpi.edu).}%

\maketitle
\begin{abstract}
Breast CT provides image volumes with isotropic resolution in high contrast, enabling detection of small calcification (down to a few hundred microns in size) and subtle density differences. Since breast is sensitive to x-ray radiation, dose reduction of breast CT is an important topic, and for this purpose, few-view scanning is a main approach. In this article, we propose a Deep Efficient End-to-end Reconstruction (DEER) network for few-view breast CT image reconstruction. The major merits of our network include high dose efficiency, excellent image quality, and low model complexity. By the design, the proposed network can learn the reconstruction process with as few as $\mathcal{O}(N)$ parameters, where $N$ is the side length of an image to be reconstructed, which represents orders of magnitude improvements relative to the state-of-the-art deep-learning-based reconstruction methods that map raw data to tomographic images directly. Also, validated on a cone-beam breast CT dataset prepared by Koning Corporation on a commercial scanner, our method demonstrates a competitive performance over the state-of-the-art reconstruction networks in terms of image quality. The source code of this paper is available at: https://github.com/HuidongXie/DEER.
\end{abstract}

Breast CT, Deep learning, Few-view CT, Low-dose CT, X-ray CT.

\section{Introduction}
\label{sec:introduction}
According to the American Cancer Society, breast cancer remains the second leading cause of cancer death among women in the United States. Approximately 40,000 people die from breast cancer each year \cite{smith_improvement_2011}. The chance of a woman having this disease during her life is 1 in 8. The wide use of x-ray mammography, which can detect breast cancer at the early stage, has helped reduce the death rate. Five-year relative survival rates by stage at diagnosis for breast cancer patients with histologic grade 1 are 97\% (local stage), 89\% (regional stage), and 24\% (distant stage) respectively \cite{henson_relationship_1991}. These data indicate that detection at an early stage plays a crucial role in significantly improving prognosis of breast cancer patients. Therefore, the development of breast imaging techniques with high performance will directly benefit these patients.

Mammography is a 2D imaging technique with structures overlapped along with the x-ray paths, severely degrading image contrast. While breast tomosynthesis is a pseudo-3D imaging technique, breast CT  provides an image volume of high quality and promises a superior diagnostic performance. Indeed, CT is one of the most essential imaging modalities extensively used in clinical practice \cite{brenner_computed_2007}. Although CT brings overwhelming healthcare benefits, it may potentially increase cancer risk due to the involved ionizing radiation \cite{lin_radiation_2010}. Since breast is particularly sensitive to x-ray radiation, dose reduction of breast CT is directly healthcare relevant. If the effective dose of routine CT examinations is reduced to 1 mSv per scan, the long-term risk of CT scans can be considered negligible. The average mean glandular dose of a typical breast CT scanner ranges between 7 and 13.9 {mSv}, while the standard radiation dose currently set by the Food and Drug Administration (FDA) is < 6 {mSv}. This gap demands major research efforts.

In the past years, several deep-learning-based low-dose CT denoising methods were proposed to reduce radiation dose with excellent results \cite{shan_3-d_2018,chen_low-dose_2017-1,shan_competitive_2019}. In parallel, few-view CT is also a promising approach to reduce the radiation dose, especially for breast CT \cite{pacile_clinical_2015} and C-arm CT \cite{floridi_c-arm_2014,orth_c-arm_2009}. Moreover, few-view CT may be implemented in mechanically stationary scanners in the future avoiding all problems associated with a rotating gantry. Recently, data-driven algorithms have shown a great promise to solve the few-view CT problem \cite{wang_perspective_2016}. For example, FBPConvNet \cite{jin_deep_2017} uses the classical U-net \cite{ronneberger_u-net_2015} structure with conveying paths to remove streak artifacts. The residual convolutional neural network (Residual-CNN) \cite{cong_deep-learning-based_2019} implements residual paths \cite{he_deep_2016} in CNN to link previous layers to later layers. Also, a shallow architecture was developed \cite{pelt_fast_2013} that learns a weighted combination of multiple filtered back-projection (FBP) \cite{wang_approximate_2007} reconstructions with different learned filters.

Few-view CT is a hot topic in the field of tomographic image reconstruction. Because of the requirement imposed by the Nyquist sampling theorem \cite{landau_sampling_1967}, reconstructing high-quality CT images from under-sampled data is traditionally considered impossible. When sufficient projection data are acquired, analytical methods such as filtered back-projection (FBP) are widely used for accurate image reconstruction. However, in the few-view CT circumstance, severe streak artifacts are introduced in analytically reconstructed images due to the incompleteness of projection data. To overcome this issue, various iterative techniques were proposed to incorporate prior knowledge in the image reconstruction process. Well-known methods include algebraic reconstruction technique (ART) \cite{gordon_algebraic_1970}, simultaneous algebraic reconstruction technique (SART) \cite{andersen_simultaneous_1984}, expectation maximization (EM) \cite{dempster_maximum_1977}, etc. Nevertheless, these iterative methods are time-consuming and still fail to produce satisfying results in many challenging cases. Recently, deep learning becomes very popular due to the development of neural network, high-performance computing (such as graphics processing unit (GPU)) and big data science and technology. In particular, deep learning has now become a new frontier of CT reconstruction research  \cite{wang_perspective_2016, wang_guest_2015,wang_image_2018}.

In the literature, only a few deep learning methods were proposed for learning a network-based reconstruction algorithm from raw data. Zhu \etal \cite{zhu_image_2018} proposed an AUTOMAP, which uses fully-connected layers to learn the mapping from raw \emph{k}-space data to the corresponding  MRI image with $\mathcal{O}(N^4)$ parameters, where $N$ denotes the side length of a reconstructed image. There is no doubt that a similar technique can be implemented to learn the mapping from the projection domain to the image domain for CT or other tomographic imaging modalities, as clearly explained in our perspective article \cite{wang_perspective_2016}. However, importing the whole sinogram into the network requires a huge amount of memory and represents a major computational challenge to train the network for a full-size CT image/volume on commercial GPUs. Moreover, using fully-connected layers to learn the mapping assumes that every single point in the data domain is related to every single point in the image domain. While this assumption is generally correct, it does not utilize the intrinsic structure of the tomographic imaging process. In the case of CT scanners, x-rays generate line integrals from different angles over a field of view (FOV) for image reconstruction. Therefore, there are various degrees of correlation between projection data within each view and at different orientations. W\"{u}rfl \etal \cite{wurfl_deep_2018} replaces the fully-connected layer in the network with a back-projection operator to reduce the computational burden. Even though their method reduces the memory cost by not storing a large matrix in the GPU, the back-projection process is no longer learnable. Hu \etal introduced a learned experts' assessment-based reconstruction network (LEARN) \cite{chen_learn:_2018} for few-view CT. LEARN maps few-view projection data to a reconstructed image through unrolling a classic iterative process in a data-driven manner. The main difference between our method and their method is that the back-projection operation in LEARN is fixed and not learnable. In contrast, our method learns an improved network-based back-projection to address the model mismatch problem encountered by analytical reconstruction methods in few-view settings and enhance reconstruction quality accordingly. Another proposed deep-learning-based CT reconstruction method \cite{li_learning_2019}, known as the iCT-Net, uses multiple small fully-connected layers and incorporates the viewing-angle information in learning the mapping from sinograms to images. iCT-Net reduces the computational complexity from $\mathcal{O}(N^4)$ for the network by Zhu \etal~to $\mathcal{O}(N^2 \times N_d)$, where $N_d$ denotes the number of CT detector elements. In most CT scanners, $N_d$ is usually equal to or greater than $N$. The complexity $\mathcal{O}(N^2 \times N_d)$ is still large for CT reconstruction. He \etal~proposed an iRadonMAP framework \cite{he_radon_2020} to simulate the inverse Radon transform by deep learning, in which a sinusoid in the filtered sinogram contributes to a single point in the image domain.

Here we propose a Deep Efficient End-to-end Reconstruction (DEER) network for low-dose few-view breast CT image reconstruction from raw measurement data. The major merits of our network include high dose efficiency, excellent image quality, and low model complexity. Computationally, the number of parameters required by DEER is as few as $\mathcal{O}(N)$. In this study, the number of parameters is set to $\mathcal{O}(N^2\times N_v)$ for better performance, where $N_v$ is the number of projections. Network that utilizes $\mathcal{O}(N)$ parameters will also be presented in Section \ref{sec::view_independent}. In the few-view CT case, $N_v$ is much less than $N_d$ which is a favorable comparison to the complexity $\mathcal{O}(N^2 \times N_d)$ of iCT-Net. Lastly, our network design allows us to split the reconstruction algorithm into several branches so that DEER can be trained to reconstruct large breast CT images ($1024\times 1024$) directly on a regular GPU. Different from the iRadonMAP method, DEER learns the reconstruction process by letting every single point in the projection domain relates to a line in the  image domain. Our proposed DEER network is more efficient than the iRadonMAP network as the DEER network does not require interpolation in the projection domain, and DEER does not need a fully-connected layer for filtering. Moreover, pretraining using another dataset is not necessary in DEER, which simplifies and speeds up the training process. Lastly, DEER can be applied to real cone-beam data obtained on a commercial scanner, while iRadonMAP is designed for parallel-beam data. In this study, real cone-beam CT data provided by Koning Corporation were used to train, validate, and test the proposed method.

The proposed DEER is inspired by the well-known filtered back-projection mechanism, and designed to learn a refined filtration and back-projection for data-driven image reconstruction. As a matter of fact, every point in the sinogram domain only relates to pixels/voxels on a single X-ray path through a FOV. This means that line integrals acquired by different detector elements at a particular angle are not directly related to each other. Also, after an appropriate filtering operation, a filtered projection profile must be smeared back over the FOV. These two ray-oriented processes suggest that the reconstruction process can be, to a large degree, learned in a point-wise manner, which is the main idea of the DEER network to reduce the memory burden. By directly comparing DEER with other recently published state-of-the-art reconstruction algorithms, based on a breast CT dataset collected by Koning Corporation on a commercial scanner, the experimental results empirically demonstrate DEER has a superior performance over existing reconstruction methods in terms of image quality.

The rest of the paper is organized as follows. In Section \ref{sec:Methodology}, the methodology is detailed. In Section \ref{sec:result}, the experimental setup is described, along with representative results and ablation study. In Section \ref{sec::discussion}, we discuss relevant issues and clinical applications of the proposed method and conclude this paper. 

\section{Methodology} \label{sec:Methodology}

\subsection{Proposed Framework}

Image reconstruction for few-view CT can be expressed as follows:
\begin{equation}
\mat{I}_{\mathrm{FullV}}=\mathcal{R}^{-1}(\mat{S}_{\mathrm{FewV}} ), \label{eqn:1}\end{equation}
where $\mat{I}_{\mathrm{FullV}}\in \mathbb{R}^{N\times N}$ is an image of $N \times N$ pixels, $\mat{S}_{\mathrm{FewV}}\in \mathbb{R}^{N_v\times N_d}$ is the sinogram of $N_v \times N_d$ data, where $N_d$ represents number of detectors, subscripts $\mathrm{FullV}$ and $\mathrm{FewV}$ stand for full-view and few-view respectively, and $\mathcal{R}^{-1}$ denotes an inverse transform \cite{barrett_iii_1984, barrett_fundamentals_1988} such as FBP in the case of sufficient 2D projection data. Alternatively, CT image reconstruction can also be transformed into a problem of solving a system of linear equations. Ideally, the FBP method produces satisfying results when sufficient high-quality projection data are available. However, when the number of linear equations is less than the number of unknown pixels/voxels in the few-view CT setting, image reconstruction becomes an undetermined problem, and even an iterative algorithm cannot reconstruct satisfactory images in difficult cases. Recently, deep learning provides a novel way to extract features of raw data for image reconstruction. With a deep neural network, training data can be utilized as strong prior knowledge to establish the relationship between a sinogram and the corresponding CT image, efficiently solving this undetermined problem. 

\begin{figure*}[!h]
\centering
\includegraphics[width=\textwidth]{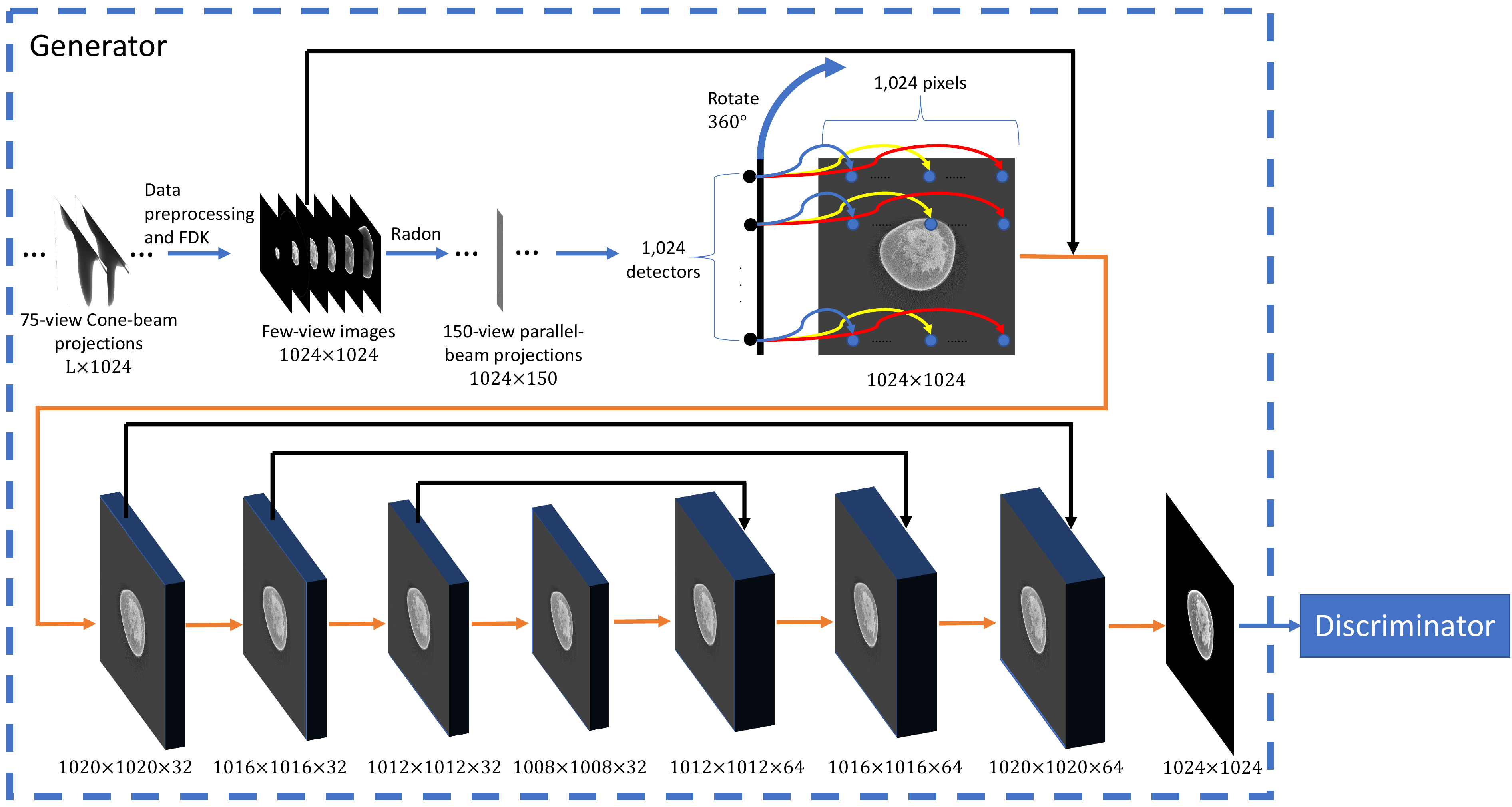}
\caption{Workflow of the proposed DEER network. The numbers below each block indicate its dimensionality. $L$ is breast length. Orange arrows indicate $5\times5$ convolutional layers with 32 filters, stride 1, and no zero-padding. Black arrows indicate concatenate operations.}
\label{fig:network_structure}
\end{figure*}

Note that in this pilot study, DEER is designed to perform 2D breast image reconstruction from 3D cone-beam data so that the memory requirement is met by our current GPU workstation. As a matter of fact, the proposed method could perform 3D CT image reconstructions because the embedded physical proprieties are the same in both 2D and 3D cases. However, graphical memory burden is still an issue for processing 3D raw data directly. The dimensionality of the raw scan provided by Koning is $1024\times300\times L$ ($1024\times N_v\times L$ in few-view cases, where $N_v$ is much less than 300. $N_v$ was selected as 75 in this study), where $L$ depends on breast size. Since the entire cone-beam dataset is needed for reconstructing a 3D image volume, there is no feasible way for us to process 3D data on our current hardware. Hence, we have developed a strategy to overcome this difficulty and still utilize the raw data from Koning. Our strategy and overall workflow of the DEER network are presented in Fig. \ref{fig:network_structure}. First, the Feldkamp, Davis, and Kress (FDK) algorithm \cite{feldkamp_practical_1984} is used to process Koning 75-view cone-beam projections (the original input) via intermediate FDK-type image reconstruction into 150-view parallel-beam projections (the intermediate output) through a pre-selected transverse slice (at a pre-specified longitudinal position). By doing so, DEER can circumvent the above-mention memory limitation and perform slice-based image reconstruction. Since parallel-beam projections are based on few-view images instead of the ground-truth image volume, the resultant parallel-beam projections are of low quality. To integrate all the information, the intermediate FDK-type few-view images are also included in the overall workflow from raw few-view data to the final image reconstruction, as illustrated in Fig. \ref{fig:network_structure}. Our ablation studies and the corresponding results will be presented in Section \ref{sec::ablation_res} to demonstrate that DEER is better than competing image-domain methods. Moreover, our previous conference publication \cite{xie_dual_2019}, which is fully based on simulations, already shows that a network-based reconstruction algorithm is better than post-processing networks in terms of image quality.

The DEER network is empowered in the Wasserstein Generative Adversarial Network \cite{arjovsky_wasserstein_2017} (WGAN) framework. Compared with the original generative adversarial network (GAN) \cite{goodfellow_generative_2014-1}, WGAN is more stable and less sensitive to the network topology. The effectiveness of the WGAN framework will also be demonstrated in Section \ref{sec::ablation_res}. In this study, the proposed framework consists of two components: a generator network $G$ and a discriminator network $D$. $G$ aims at reconstructing images directly from a batch of few-view sinograms. $D$ receives images from either $G$ or the ground-truth dataset, and intends to distinguish whether the input image is real (the ground-truth) or fake (from $G$). Both networks can optimize themselves in the training process. If an optimized network $D$ can hardly distinguish fake images from real ones, then we say that generator $G$ can fool discriminator $D$, which is the goal of WGAN. By design, the network $D$ also helps improve the texture of the final image and prevent over-smoothing from occurring.

Different from the vanilla GAN \cite{goodfellow_generative_2014-1}, WGAN replaces the cross-entropy loss function with the Wasserstein distance, improving the training stability. In the WGAN framework, the 1-Lipschitz function is assumed with weight clipping. However, it was pointed out \cite{gulrajani_improved_2017} that weight clipping may be problematic in WGAN, and it can be replaced with a gradient penalty, which is implemented in our proposed framework. Hence, the objective function of the network $D$ is expressed as follows:

\begin{align}
\underset{\mat{\theta}_G}{\max}\ \underset{\mat{\theta}_D}{\min}&\underbrace{\bigg\{\mathbb{E}_{\mat{S}_{\mathrm{FewV}}}\left[D(G(\mat{S}_{\mathrm{FewV}}))\right]-\mathbb{E}_{\mat{I}_{\mathrm{FullV}}}\left[D(\mat{I}_{\mathrm{FullV}})\right]}_{\mathrm{Wasserstein\  distance}}\notag\\ 
&+\lambda\underbrace{\mathbb{E}_{\bar{\mat{I}}}\left[(\|\nabla(\bar{\mat{I}})\|_2-1)^2\right]}_{\mathrm{gradient\  penalty}}\bigg\} \label{eqn:2}
\end{align}
where $\mat{S}_{\mathrm{FewV}}$ and $\mat{I}_{\mathrm{FullV}}$ represent few-view sinograms and ground-truth images respectively, $\mathbb{E}_a[b]$ denotes the expectation of b as a function of a, $\mat{\theta}_G$ and $\mat{\theta}_D$ represent the trainable parameters of networks $G$ and $D$ respectively, $\bar{\mat{I}}=\alpha\cdot \mat{I}_{\mathrm{FullV}}+(1-\alpha)\cdot G(\mat{S}_{\mathrm{FewV}})$, and $\alpha$ is uniformly sampled from the interval [0,1]. In other words, $\bar{\mat{I}}$ represents images between fake and real images. $\nabla(\bar{\mat{I}})$ denotes the gradient of $D$ with respect to $\bar{\mat{I}}$, and $\lambda$ is a parameter used to balance the Wasserstein distance term and gradient penalty term. As suggested in \cite{goodfellow_generative_2014-1, arjovsky_wasserstein_2017, gulrajani_improved_2017}, the networks $D$ and $G$ are updated alternatively.

\subsection{Generator Network}

The input to DEER is the original 75-view cone-beam data. First, a normalized logarithm operation is taken to convert the raw intensity profiles to line integrals in terms of linear attenuation coefficients along with other data preprocessing steps suggested by Koning. Further data preprocessing steps include applying Shepp-Logan filter and excluding abnormal detector measurements. The FDK few-view reconstructions are then performed to offer a basis for estimating parallel beam projections through a pre-specified transverse image slice. After that, 150-view parallel beam projections are produced via ray-tracing at an adequate sampling rate, and further filtered using the Shepp-Logan kernel, which is the suggested filter choice by Koning. The reason why we only estimated 150-view parallel-beam projections from the FDK-reconstructed volumes is that 150-view is the largest number of views that our current workstation can handle in an acceptable amount of training time. The conversion from divergent beam data to parallel beam data is to speed up the training process within the available GPU memory (otherwise, we need to load the entire cone-beam data volume into the GPU memory). In DEER, the filtration is performed in the Fourier domain via multiplication. Then, the filtered sinogram data are passed into the generator network $G$. The network $G$ learns a network-based back-projection and outputs a reconstructed image. This generator network can be divided into two components: back-projection and refinement.

First, the filtered sinogram data are passed into the back-projection part of the network $G$. This part aims at reconstructing a batch of images from projection data. As illustrated in Fig. \ref{fig:network_structure}, the reconstruction algorithm is inspired by the following intuition: every point in the sinograms only relates to pixel values on the associated x-ray path through the underlying image, and other pixels contribute little to it. With this intuition, the reconstruction process is learned in a point-wise manner using a point-wise fully-connected layer, and DEER can truly learn the back-projection process with a computational complexity of $\mathcal{O}(N)$ parameters, thereby reducing the memory overhead. Put differently, for a parallel-beam sinogram with dimensionality of $N_v\times N$, there is a total of $N_v\times N$ small fully-connected layers in the proposed network. The input to each of these small fully-connected layers is a single point in the sinogram domain, and the output is a line-specific vector of $N\times 1$ elements. After this point-wise fully-connected layer, rotation and summation are applied to simulate the FBP method, putting all the learned lines back to where they should be. Bilinear interpolations \cite{gribbon_novel_2004} are used to keep the rotated images on a Cartesian grid. This network design allows the neural network to learn the reconstruction process with only $N$ parameters if all the small fully-connected layers share the same weight. However, due to the complexity of medical images, incomplete projection data, and involved bilinear interpolations, $N$ parameters are not sufficient to produce high-quality images. Therefore, we increased the number of parameters to $\mathcal{O}(N^2\times N_v)$ for this point-wise learning network. That is, we made network parameters specific to both viewing angles and X-ray paths to compensate for artifacts from bilinear interpolations and other factors. There is no bias term in the back-projection process.

Images reconstructed in the back-projection portion are concatenated with the intermediate few-view FDK images and then feed into the refinement portion of the network $G$. Although the proposed filtration and back-projection parts learn a refined FBP method, streak artifacts cannot be completely eliminated. The refinement part of $G$ is used to remove the remaining artifacts. It is a 9-layer U-net \cite{ronneberger_u-net_2015}. U-net was originally designed for image segmentation and has been utilized in various medical imaging applications. For example, \cite{shan_3-d_2018,chen_low-dose_2017} use U-net with conveying paths for CT images denoising, \cite{jin_deep_2017,lee_deep-neural-network-based_2019} for few-view CT, and \cite{quan_compressed_2018} for sparse data MRI \cite{donoho_compressed_2006}. Each layer in the proposed U-net is followed by a rectified linear unit (ReLU). 32 kernels of $5\times 5$ are used in both convolutional and transpose-convolutional layers in the U-net. A stride of 1 is used for all down-sampling and up-sampling layers. Zero-padding is not implemented in the U-net. Conveying paths are implemented to reserve high-level information and improve reconstruction quality.

\subsection{Discriminator Network}
The discriminator network $D$ takes an image from either $G$ and the ground-truth dataset, trying to distinguish whether the input is real or fake. The discriminator network has 6 convolutional layers with 64, 64, 128, 128, 256, 256 filters respectively, which are followed by 2 fully-connected layers with the number of neurons 1024 and 1 respectively. The leaky ReLU activation function is enforced after each layer with a slope of 0.2 in the negative part. Convolution operations are performed with $3\times3$ windowing and zero-padding for all convolutional layers. Stride equals to 1 for odd layers and 2 for even layers. 

\subsection{Objective Functions for the Generator Network}

The objective function used to optimize the network $G$ involves mean absolute error (MAE) \cite{willmott_advantages_2005} and structural similarity index (SSIM) \cite{zhou_wang_image_2004}. Compared with the popular mean square error (MSE) \cite{chen_low-dose_2017,wolterink_generative_2017}, which is also a mean-based measurement, MAE does not over-penalize significant differences and does not tolerate subtle errors in the reconstructed images. Therefore, MAE addresses some disadvantages of the MSE loss such as the over-smoothed issue. The formula of the MAE loss is expressed as follows:
\begin{equation}
\mathcal{L}_1=\frac{1}{N_b W  H}\sum_{i=1}^{N_b}\left|\mat{Y}_i-\mat{X}_i\right|,  \label{eqn:3}
\end{equation}
where $N_b$, $W$ and $H$ denote the number of batches, the width and height of involved images, $\mat{Y}_i$ and $\mat{X}_i$ represent the ground-truth images and images reconstructed by the network $G$ respectively. 

To ensure that the back-projection portion of the generator network $G$ can learn an accurate reconstruction from projection data. MAE between the images reconstructed by the back-projection part and the ground-truth images is also added as part of the objective function. The formula is expressed as follows:
\begin{equation}
\mathcal{L}_1^{\mathrm{bp}}=\frac{1}{N_b W  H}\sum_{i=1}^{N_b}\left|\mat{Y}_i-\mat{X}_{i}^{\mathrm{bp}}\right|,  \label{eqn:4}
\end{equation}
where $\mat{X}_{i}^{bp}$ denotes the image reconstructed by the back-projection portion.

Since MAE does not address the inter-dependencies between pixels in medical images, SSIM is introduced in the objective function to acquire visually appealing results. SSIM measures structural similarity between two images. The convolution window used to measure SSIM is set to $11\times 11$. The SSIM formula is expressed as follows:
\begin{equation}
\mathrm{SSIM}(\mat{Y}, \mat{X})=\frac{(2\mu_{\mat{Y}}\mu_{\mat{X}}+C_1)(2\sigma_{\mat{YX}}+C_2)}{(\mu_{\mat{Y}}^2+\mu_{\mat{X}}^2+C_1)(\sigma_{\mat{Y}}^2+\sigma_{\mat{X}}^2+C_2)},  \label{eqn:5}
\end{equation}
where $C_1=(K_1\cdot R)^2$ and $C_2=(K_2\cdot R)^2$ are constants to stabilize the ratios when the denominator is too small, $R$ stands for the dynamic range of pixel values, often times $K_1=0.01$, $K_2=0.03$, $\mu_{\mat{Y}}$, $\mu_{\mat{X}}$, ${\sigma_{\mat{Y}}}^2$, ${\sigma_{\mat{X}}}^2$ and $\sigma_{\mat{YX}}$ are the means of $\mat{Y}$ and $\mat{X}$, deviations of $\mat{Y}$ and $\mat{X}$, and the correlation between $\mat{Y}$ and $\mat{X}$ respectively. Then, the structural loss becomes the following: 
\begin{equation}
\mathcal{L}_{\mathrm{sl}}=1-\mathrm{SSIM}(\mat{Y},\mat{X}). \label{eqn:6}
\end{equation}
The adversarial loss helps the generator network produce faithful images that are indistinguishable by the discriminator network. In reference to Eq.~\eqref{eqn:2}, the adversarial loss is expressed as follows:
\begin{equation}
\mathcal{L}_{\mathrm{al}}=-\mathbb{E}_{\mat{S}_{\mathrm{FewV}}}[D(G(\mat{S}_{\mathrm{FewV}}))]. \label{eqn:7}
\end{equation}

The overall objective function of $G$ is then summarized as follows:
\begin{equation}
\mathcal{L}_G=\lambda_{\mathrm{al}} \mathcal{L}_{\mathrm{al}}+\lambda_{\mathrm{sl}} \mathcal{L}_{\mathrm{sl}} + \mathcal{L}_1 + \mathcal{L}_1^{\mathrm{bp}}, \label{eqn:8}
\end{equation}
where $\lambda_{\mathrm{al}}$ and $\lambda_{\mathrm{sl}}$ are hyper-parameters to balance different loss components.

\subsection{Parameter Settings and Training Details}

The hyper-parameter selection is an important issue. To balance the Wasserstein distance and the gradient penalty, the hyperparameter $\lambda$ was set to 10, as suggested in the original paper \cite{gulrajani_improved_2017}, while $\lambda_{\mathrm{al}}=0.0025$, and $\lambda_{\mathrm{sl}}=0.8$ were experimentally obtained on the validation dataset. These hyperparameters were adjusted for the best SSIM, since it is a widely preferred indicator over MSE and MAE for image quality. All codes were implemented on the TensorFlow platform \cite{45166} using an NVIDIA Titan RTX GPU with a graphical memory of 24 gigabytes. The Adam optimizer optimized the parameters \cite{DBLP:journals/corr/KingmaB14} with $\beta_1=0.9$ and $\beta_2=0.999$.

Network training can be divided into two phases. In the first phase, the back-projection part of the network was first pre-trained for 10 epochs. In the second phase, the network was trained as a whole, but the learning rate for the parameters in the back-projection portion was ten times smaller than the learning rate for the other parts of the network. A batch size of 5 was used for training in the first training phase and 3 in the second training phase.

\begin{figure*}[!h]
\centering
\includegraphics[width=\textwidth]{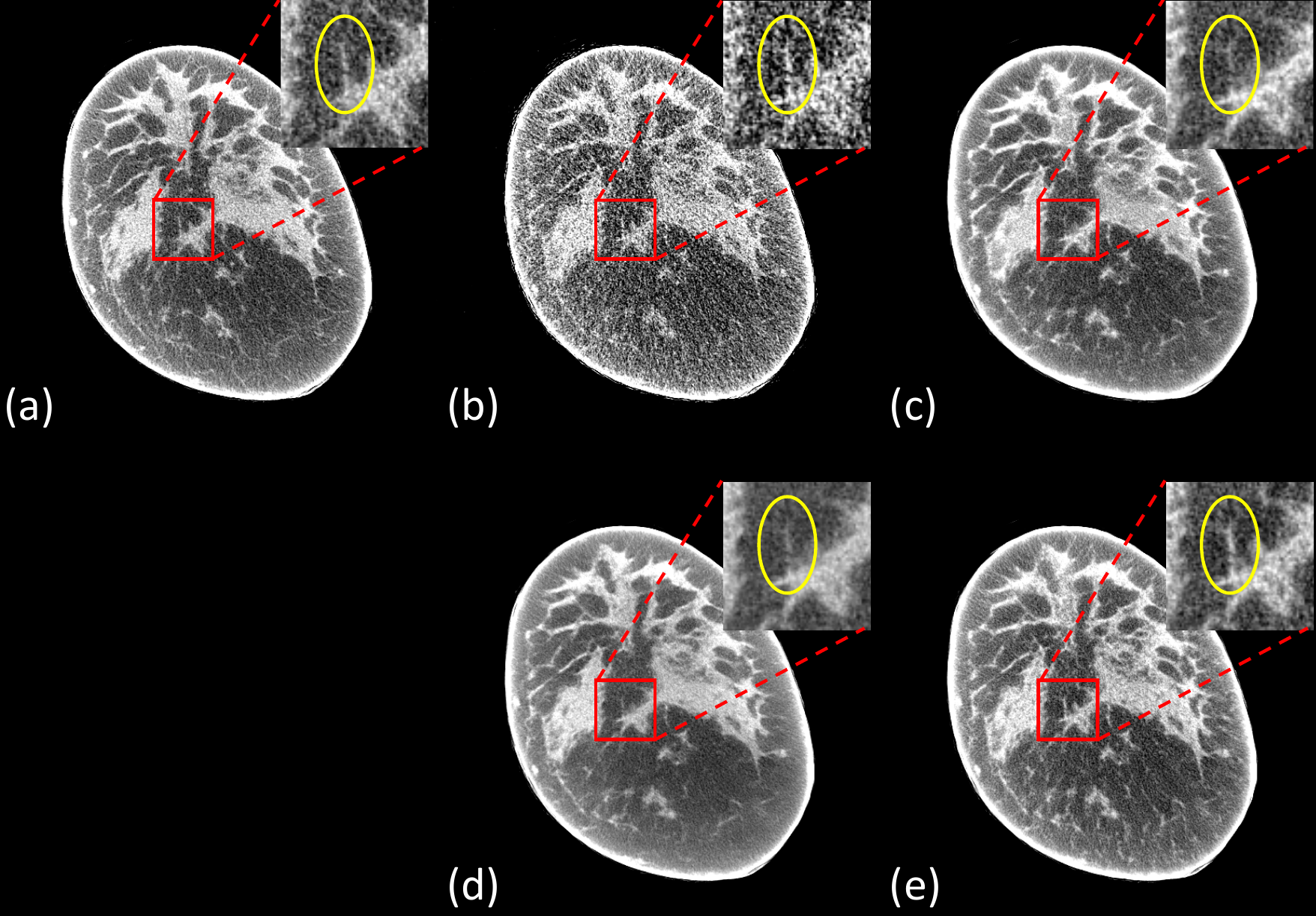}
\caption{Representative slices reconstructed using different methods for Koning dataset. (a) Ground-truth, (b) FDK from 75-view cone-beam data, (c) Residual-CNN, (d) FBPConvNet, (e) DEER. The red boxes mark the Regions of Interest (ROIs). Yellow circles mark some subtle details in the ROIs. The display window is [-200, 200] HU for visualizing breast details. Note that the final reconstructions were post-processed to remove irrelevant structures outside the field of view.}
\label{fig:result1}
\end{figure*}

\begin{figure*}[!h]
\centering
\includegraphics[width=\textwidth]{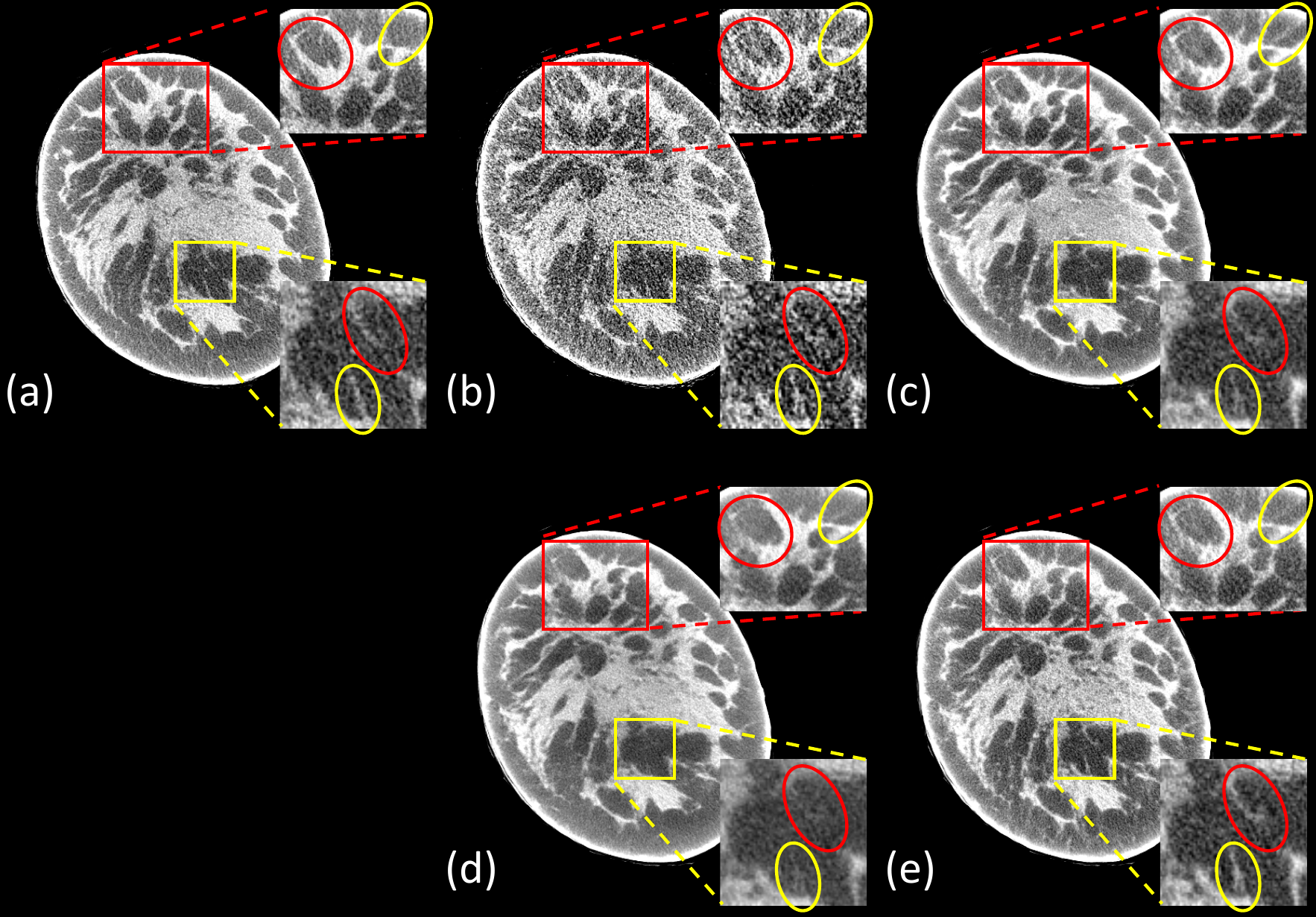}
\caption{Representative slices reconstructed using different methods for Koning dataset. (a) Ground-truth, (b) FDK from 75-view cone-beam data, (c) Residual-CNN, (d) FBPConvNet, (e) DEER. The red and yellow boxes mark the Regions of Interest (ROIs). Red and yellow circles mark some subtle details in the ROIs. The display window is [-200, 200] HU. Note that the final reconstructions were post-processed to remove irrelevant structures outside the field of view.}
\label{fig:result2}
\end{figure*}

\section{Results} \label{sec:result}

\subsection{Data and Experimental Design}

A clinical female breast dataset was used to train the proposed DEER method and evaluate its performance. The dataset was generated and prepared by Koning Corporation. These breast images were acquired on a state-of-the-art breast CT scanner designed and manufactured by Koning. Totally, 19,575 breast CT images were acquired from 42 patients. All the images were reconstructed from 300 cone-beam projections at 42 peak kilovoltage (kVp), which were used as the ground-truth images to train the proposed network. The distance between the x-ray source and the patient is 650 millimeters, while the distance between the patient and the detector array is 273 millimeters. All the images are of $1024 \times 1024$ (four times larger than clinical CT slices). Totally, 32 patients (13,970 images) were randomly selected for training, 3 patients (1,197 images) for validation, and 7 patients (4,408 images) for testing. In the training, validation, and testing phases, 75-view cone-beam projections provided by Koning were used as the input to the DEER network.

\begin{table*}[htbp]
\centering
\caption{Quantitative assessments on different methods ($\mathrm{MEAN}\pm \mathrm{STD}$) for 75-view reconstructions. For each metric, the best result is marked in bold face. The measurements were obtained by averaging the values in the testing set.}
\begin{tabular}{ccccc}
\toprule
&                         FDK &      FBPConvNet       & Residual-CNN          & DEER \\
\midrule
PSNR & $27.5738\pm8.9023$   & \pmb{$32.9347\pm6.8931$}   & $32.6871\pm9.3807$   & $31.9676\pm6.0831$\\
SSIM & { }{ }$0.8988\pm0.0978$    & { }{ }$0.9088\pm0.0679$    & { }{ }$0.9363\pm0.0651$    & { }{ }\pmb{$0.9372\pm0.0644$}\\
MAE  & { }{ }$0.0158\pm0.0166$    & { }{ }$0.0099\pm0.0111$    & { }{ }$0.0091\pm0.0111$    & { }{ }\pmb{$0.0087\pm0.0108$}\\
\bottomrule
\end{tabular}
\label{table:quan_koning}
\end{table*}

For qualitative assessment, we compared DEER with two state-of-the-art deep learning methods, including FBPConvNet \cite{jin_deep_2017} and Residual-CNN \cite{cong_deep-learning-based_2019}. Both of these methods are considered as image-domain methods, in which the FBP operation is fixed and not learnable. There was no need for parameter fine-tuning since each method only used one training loss. The Euclidean loss was used in FBPConvNet, and the structural similarity index (SSIM) was used in Residual-CNN. Lastly, images reconstructed by the FDK algorithm from 75-view cone-beam raw data were also added as a baseline for comparison. Raw 75-view equidistant cone-beam projections were selected from the 300-view cone-beam projections for FDK reconstructions. Since we were not able to train iCT-Net and AUTOMAP for $1024\times 1024$ image reconstruction on our GPU workstation due to their huge memory requirements (in the case of $1024\times 1024$ image reconstruction, instead of $512\times 512$ or even smaller-scale image reconstruction), these two methods are not included for assessment in this study.

\begin{figure*}[!t]
\centering
\includegraphics[width=\textwidth]{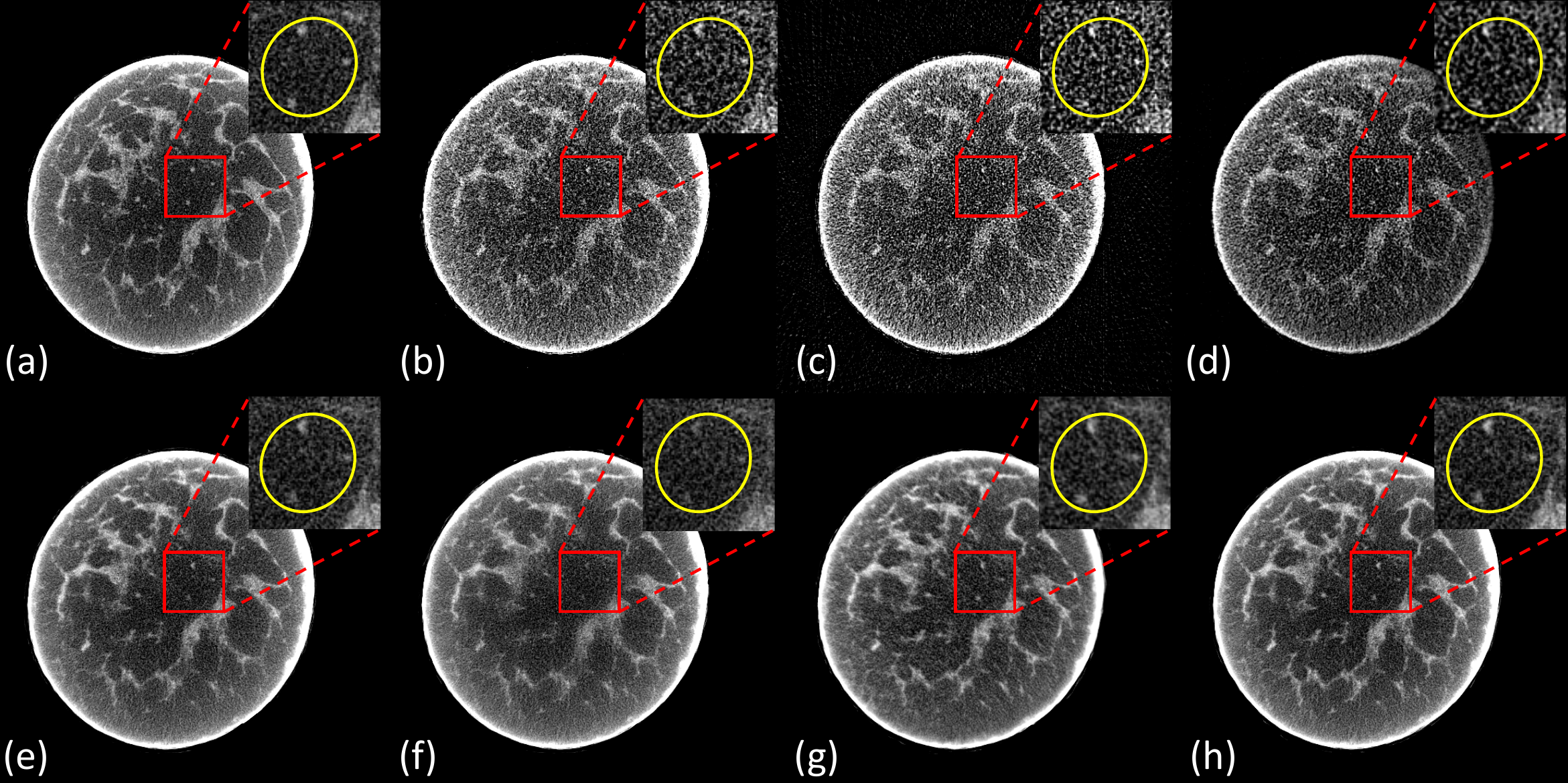}
\caption{Representative slices reconstructed using different methods. (a) The ground-truth, (b) FDK reconstruction from 75-view cone-beam projections, (c) FBP from 150 simulated parallel-beam projections, (d) DEER-BP, (e) DEER-NoWGAN, (f) DEER-FBP, (g) DEER-Sino, and (h) DEER. The red boxes mark the Regions of Interest (ROIs). Yellow circles mark some subtle details in the ROIs. Note that the final reconstructions were post-processed to remove irrelevant structures outside the field of view.}
\label{fig:ablation_res}
\end{figure*}

\subsection{Comparison With Other Deep-learning Methods}

To visualize the performance of different methods, a few representative slices were selected from the testing dataset. Figs. \ref{fig:result1} and \ref{fig:result2} show results reconstructed using different methods from 75-view cone-beam projections. For better evaluation of the image quality, the regions of interest (ROIs) marked in the red/yellow boxes in both figures are magnified. Three metrics, including Peak Signal to Noise Ratio (PSNR) \cite{korhonen_peak_2012}, SSIM \cite{zhou_wang_image_2004}, and MAE \cite{willmott_advantages_2005}, were computed for quantitative assessment. The quantitative results are shown in Table \ref{table:quan_koning}.

Our proposed DEER network produced few-view denoising results comparable or superior to that from both image-domain methods. All of the deep-learning methods could effectively remove streak artifacts introduced by the few-view constraint. Particularly, the DEER network produces better reconstructions in the selected ROIs. Both image-domain methods tend to smooth out some subtle details embedded in the noisy background due to insufficient data. For example, in Fig. \ref{fig:result1}, the breast feature inside the yellow circle is hardly distinguishable in Fig. \ref{fig:result1} (c) and (d). Also, in Fig. \ref{fig:result2}, the feature in the red circle of red ROI and the features in the yellow ROI are hardly visible in the image reconstructed by FBPConvNet.

For the Residual-CNN method, the intensity of certain features is dimmer than expected in the reconstructed breast slices (feature in the yellow circle of Fig. \ref{fig:result1}, and features in the yellow ROI of Fig. \ref{fig:result2}). Moreover, the FBP method is associated with artifacts that do not exist in the ground-truth images, and in some cases, the image-domain methods cannot remove these artifacts through convolutional operations. For example, the artifact in the yellow circle of red ROI in Fig. \ref{fig:result2} is clearly visible in the image produced by Residual-CNN, but this artifact is mostly removed by DEER through a learnable network-based reconstruction algorithm.

For the quantitative assessments, DEER had better SSIM and MAE values and a slightly lower PSNR value than FBPConvNet. FBPConvNet achieved the best PSNR value due to the implementation of the Mean Squared Error (MSE) based objective function. However, the literature has discussed that higher PSNR values do not guarantee the denoising performance, especially the textural/visual similarities relative to the ground-truth images~\cite{yang_low-dose_2018,shan_3-d_2018}. Also, it should be noted that since both FBPConvNet and residual-CNN only use a single loss function for optimization, these two methods may be subject to potential losses in visual performance. Both loss functions have their own limitations, and one should not solely rely on them for estimating image quality \cite{hore_is_2013, hore_image_2010}. Even though DEER does not achieve significant improvements quantitatively, images reconstructed by DEER present promising visual comparisons. Moreover, as presented in Figs.~\ref{fig:result1} and~\ref{fig:result2}, the images reconstructed by FBPConvNet appear over-smoothed with less visual image texture, which is not desirable in clinical diagnosis. Lastly, the implementation of WGAN framework may negatively affect the quantitative measurements but it provides better recovery of subtle details and structural features \cite{shan_3-d_2018, ledig_photo-realistic_2017}. Compared with the other deep learning methods, DEER demonstrates a competitive performance in removing artifacts and reserving subtle but vital details compared with the other methods. In terms of reconstruction time, DEER takes about 0.1422 seconds to reconstruct a single 2D slice ($1024\times1024$) on an NVIDIA Titan RTX GPU.

\subsection{Ablation Studies and Other Relevant Experiments} \label{sec::ablation_res}

To demonstrate the effectiveness of the proposed method, three additional networks were trained and compared with the DEER network. In the first network, few-view images reconstructed by the FDK algorithm are eliminated from the DEER network, and the 150-view parallel-beam projections become the only input to the network. In the second network, few-view images reconstructed by the FDK algorithm are used as the only input to the network, and the back-projection part of the generator network $G$ is eliminated. Lastly, another DEER network without the WGAN component is trained and included for comparison to demonstrate the effectiveness of the WGAN framework in this task. The first network is denoted as DEER-Sino, the second network is denoted as DEER-FBP, and the third network is denoted as DEER-NoWGAN. Note that in the DEER network, the dimensionality of the input to the refinement portion of the generator network $G$ is $N_b\times1024\times1024\times2$. For a fair comparison, two copies of the input image were concatenated as the input to the DEER-Sino and DEER-FBP networks (i.e., in DEER-Sino, two copies of the image reconstructed by the back-projection part are concatenated as the input to the refinement part of the network; in DEER-FBP, two copies of the image reconstructed by the FDK algorithm are concatenated as the input to the refinement part of the network). The parameters and hyperparameters were fine-tuned for all three networks. In DEER-Sino, $\lambda_{\mathrm{al}}=0.002$, and $\lambda_{\mathrm{sl}}=0.65$. In DEER-FBP, $\lambda_{\mathrm{al}}=0.0025$, and $\lambda_{\mathrm{sl}}=0.65$. In DEER-NoWGAN, $\lambda_{\mathrm{sl}}=0.8$.
\begin{table*}[!h]
\centering
\caption{Quantitative assessment on different methods ($\mathrm{MEAN}\pm\mathrm{STD}$) for 75-view reconstructions. For each metric, the best result is marked in bold face. The measurements were obtained by averaging the values on the testing dataset. Significant streak artifacts outside the FOV were introduced in the Radon/iRadon operations, leading to poor values of the FBP method.}
\resizebox{\textwidth}{!}{
\begin{tabular}{cccccccc}
\toprule
     &  FDK         &   FBP  &        DEER-BP     &  DEER-NoWGAN &  DEER-FBP    &      DEER-Sino         & DEER \\
\midrule
PSNR & $27.5738\pm8.9023$ & $22.8804\pm7.5900$ & $25.4696\pm8.5661$ & $31.8683\pm5.9122$ &$31.7790\pm5.8244 $ & $29.0397\pm7.2356$   & \pmb{$31.9676\pm 6.0831$}\\
SSIM & { }{ }$0.8988\pm 0.0978$ & { }{ }$0.3235\pm0.1577$  & { }{ }$0.8042\pm0.1000$  & { }{ }$0.9365\pm0.0649$  &{ }{ }$0.9360\pm0.0656$  & { }{ }$0.9013\pm0.1041$    & { }{ }\pmb{$0.9372\pm0.0644$}\\
MAE  & { }{ }$0.0158\pm0.0166$ & { }{ }$0.0464\pm0.0282$  & { }{ }$0.0242\pm0.0234$  & { }{ }$0.0088\pm0.0110$  & { }{ }$0.0088\pm0.0109 $  & { }{ }$0.0132\pm0.0161 $    & { }{ }\pmb{$0.0087\pm0.0108$}\\
\bottomrule
\end{tabular}
}
\label{table:quan_ablation}
\end{table*}

As presented in Fig. \ref{fig:ablation_res}, a representative slice is selected to visualize the performance of different methods. The image reconstructed from simulated 150-view parallel-beam projections using FBP, and the output from the back-projection portion in DEER are also included for comparison, which are denoted as FBP and DEER-BP respectively. Corresponding quantitative measurements are shown in Table \ref{table:quan_ablation}. Note that the image quality is well indicated by the three white features in the yellow circle in the zoomed-in areas of Fig. \ref{fig:ablation_res}. Clearly, relatively simple post-processing methods tend to smooth-out or ignore these subtle but crucial details in the reconstructed images. However, a network-based reconstruction algorithm is better at recovering these details buried in the few-view artifacts. Although DEER-Sino has worse quantitative measurements than that of DEER-FBP, it accurately recovers these subtle features, which is desirable for clinical studies. It should be noted that the quality of the input to the DEER-Sino network is much lower than that of the DEER-FBP. Lastly, while results reconstructed by DEER-NoWGAN and DEER have similar quantitative values, certain features in images produced by DEER-NoWGAN appear to be dimmer than expected (e.g., the white feature at the bottom of the yellow circle in Fig. \ref{fig:ablation_res}). 

\begin{figure}[!h]
\centering
\includegraphics[width=0.45\textwidth]{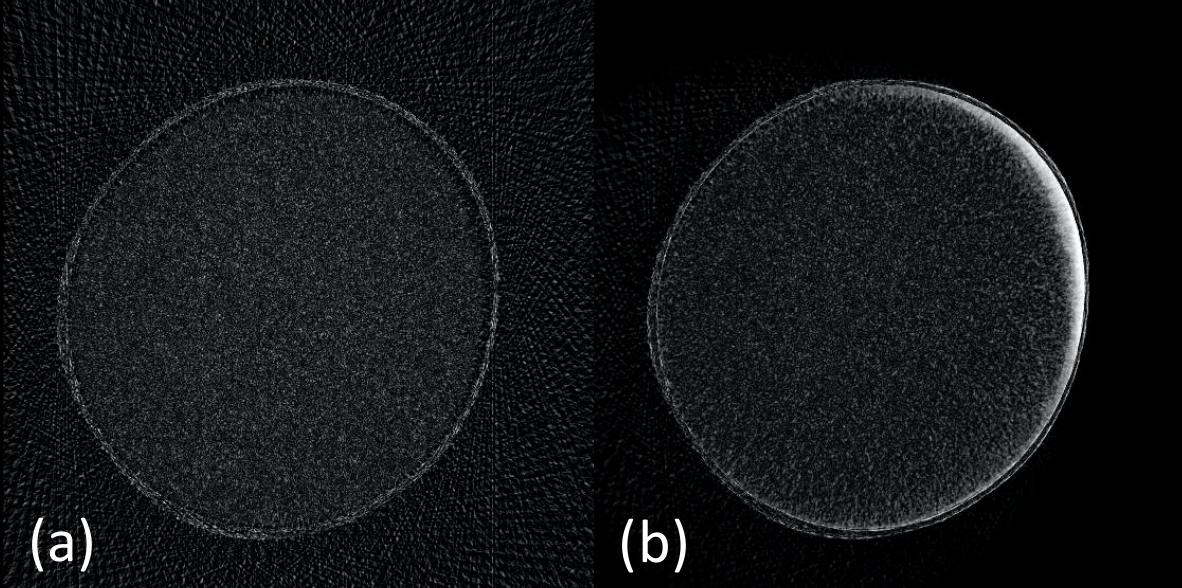}
\caption{Absolute noise maps generated from images reconstructed by different methods, referenced to the ground-truth image. (a) FBP from 150-view simulated data, (b) DEER-BP.}
\label{fig:noise_map}
\end{figure}

In few-view cases, analytical reconstruction methods such as FBP and FDK cannot produce high-quality results. The reconstructed results from post-processing image-based method, DEER-FBP, presented in Fig. \ref{fig:ablation_res} (f), cannot separate true features and artifacts. The main intuition of DEER is to learn a network-based reconstruction algorithm and to address the disadvantages of analytical reconstruction methods. As shown in the ROI of Fig. \ref{fig:ablation_res} (d) and Fig. \ref{fig:noise_map}, the proposed network-based reconstruction method helps reduce the amplitude of noise, compared with the FBP result presented in Fig. \ref{fig:ablation_res} (c). Then, the real features are more visible to the network. Note that as shown in Fig. \ref{fig:ablation_res} (d) and Fig. \ref{fig:noise_map}, DEER-BP appears to be less sensitive to the breast boundaries, resulting in darker boundaries compared with the image reconstructed by analytical FBP method. However, DEER-BP is an intermediate output from the proposed method, and the issue can be effectively addressed by the refinement portion of the network. As presented in Fig. \ref{fig:ablation_res} (g), the DEER-Sino network, in which the input is only the projection data, does not produce images with dark boundaries. Lastly, compared with FBPConvNet, residual-CNN, and DEER-NoWGAN, the WGAN framework was implemented in DEER for better recovery of subtle details and structural features but may result in a cost of compromising the quantitative measurements such as SSIM, and PSNR \cite{shan_3-d_2018}.

It is clearly shown in Fig. \ref{fig:ablation_res} that the back-projection portion can learn a network-based reconstruction that is similar to the classic FBP method. The learned FBP allows the refinement part to recover some subtle features lost in image-domain networks. As presented in Fig. \ref{fig:noise_map}, the learned back-projection produces images with overall lower noise contents than the counterparts reconstructed by FBP, and removes most of the streak artifacts outside the FOV.

\begin{figure*}[!ht]
\centering
\includegraphics[width=\textwidth]{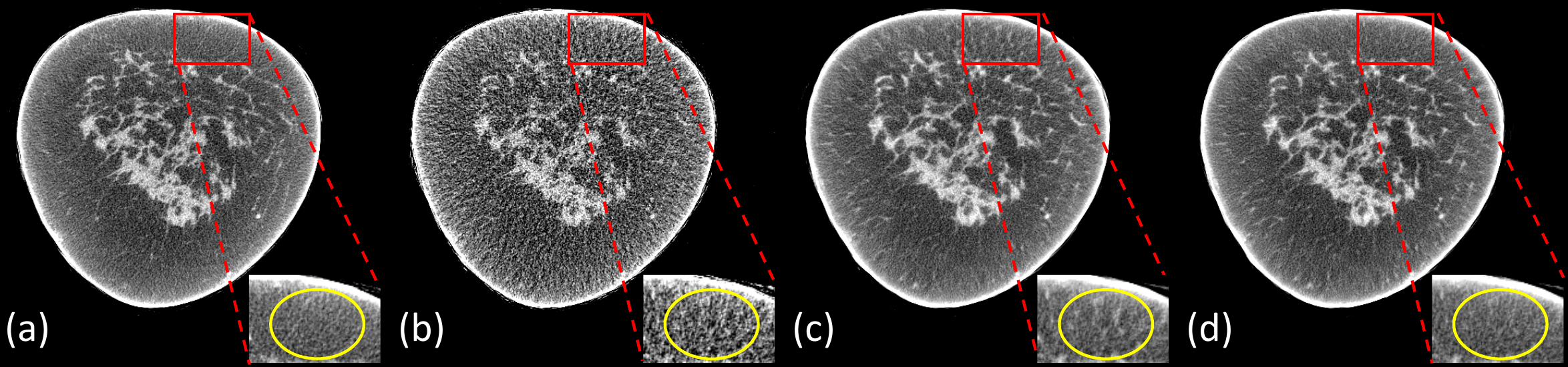}
\caption{Representative slice reconstructed using different methods for Koning dataset. (a) Ground-truth, (b) FDK from 75-view cone-beam data, (c) DEER-Lite, (d) DEER. The red boxes mark the Regions of Interest (ROIs). Yellow circles mark some subtle details in the ROIs. The display window is [-200, 200] HU. Note that the final reconstructions were post-processed to remove irrelevant structures outside the field of view.}
\label{fig:ON_result}
\end{figure*}

\subsection{View-independent Network} \label{sec::view_independent}

As described in Section \ref{sec:Methodology}, DEER utilized $\mathcal{O}(N^2\times N_v)$ to improve quality of reconstructed images. However, the proposed ray-tracing idea could be trained with as few as $\mathcal{O}(N)$ parameters. In this subsection, another DEER network (denoted as DEER-Lite) was built with $\mathcal{O}(N)$ parameters by sharing all the parameters in the reconstruction part. To be specific, in DEER, projections at different angles are back-projected to the FOV using different sets of trainable parameters (e.g., projection data at $0^{\circ}$ are back-projected using trainable parameters at $0^{\circ}$, and projection data at $90^{\circ}$ are back-projected using trainable parameters at $90^{\circ}$). On the other hand, in DEER-Lite, projections at different angles are back-projected to the FOV using the same set of parameters. The amount of trainable parameters is reduced to $\mathcal{O}(N^2)$ by sharing the parameters at different view angles. To further reduce the number of trainable parameters to $\mathcal{O}(N)$, parameters in DEER-Lite are also shared between all ray-tracing lines (i.e., all the points in the sinograms use the same set of parameters for back-projecting over the FOV). DEER is classified as a view-dependent network since different sets of parameters are used for different angles. DEER-Lite is a view-independent network since all projection angles use the same set of parameters.

DEER-Lite was trained in the same way as described in Section \ref{sec:Methodology}, except DEER-Lite uses $\mathcal{O}(N)$ parameters, while DEER uses $\mathcal{O}(N^2\times N_v)$ parameters in the back-projection part. As shown in Fig. \ref{fig:ON_result}, a representative slice was selected to present the difference in reconstruction quality of DEER and DEER-Lite. The corresponding quantitative assessments are shown in Table~\ref{table:ON_quan}. Both networks demonstrate superior denoising performance. Even though DEER only shows slight improvements in terms of selected quantitative measurements, it demonstrates more stable and clinically-favorable reconstructions in certain testing patients. For example, in the red ROI of Fig. \ref{fig:ON_result}, DEER-Lite produces some fake features that do not exist in both ground-truth images and results produced by DEER. Moreover, these unreal features are visible throughout this image volume. One possible reason for this phenomenon in DEER-Lite is that, since the proposed network-based reconstruction algorithm is learning a revised back-projection mechanism, it learns to compensate for some noise and artifacts during the reconstruction. And because the parameters are shared in all views, these compensations may produce artifacts in other views. On the other hand, DEER does not have this problem because the parameters are view-dependent in the reconstruction part, and compensations at a specific angle will not affect back-projection at another angle. Lastly, since the back-projected profiles are rotated to the corresponding angles during reconstructions, and the involved interpolations can negatively affect the image quality, a view-dependent network provides more degrees of freedom to address this issue by learning different ray-tracing patterns using different sets of parameters.

\begin{table}[!h]
\centering
\caption{Quantitative assessment on DEER and DEER-Lite ($\mathrm{MEAN}\pm \mathrm{STD}$) for 75-view reconstructions. For each metric, the best result is marked in bold face. The measurements were obtained by averaging the values on the testing dataset. }
\resizebox{0.45\textwidth}{!}{
\begin{tabular}{cccc}
\toprule
     &  FDK     &      DEER-Lite         & DEER \\
\midrule
PSNR & $27.5738\pm8.9023$ & $31.7925\pm5.8961$ & \pmb{$31.9679\pm6.0831$} \\
SSIM & { }{ }$0.8988\pm 0.0978$ & { }{ }$0.9367\pm0.0695$  & { }{ }\pmb{$0.9372\pm0.0644$}  \\
MAE  & { }{ }$0.0158\pm0.0166$ & { }{ }$0.0089\pm0.0109$  & { }{ }\pmb{$0.0087\pm0.0108$} \\
\bottomrule
\end{tabular}
}
\label{table:ON_quan}
\end{table}

\begin{figure}[!h]
\centering
\includegraphics[width=0.45\textwidth]{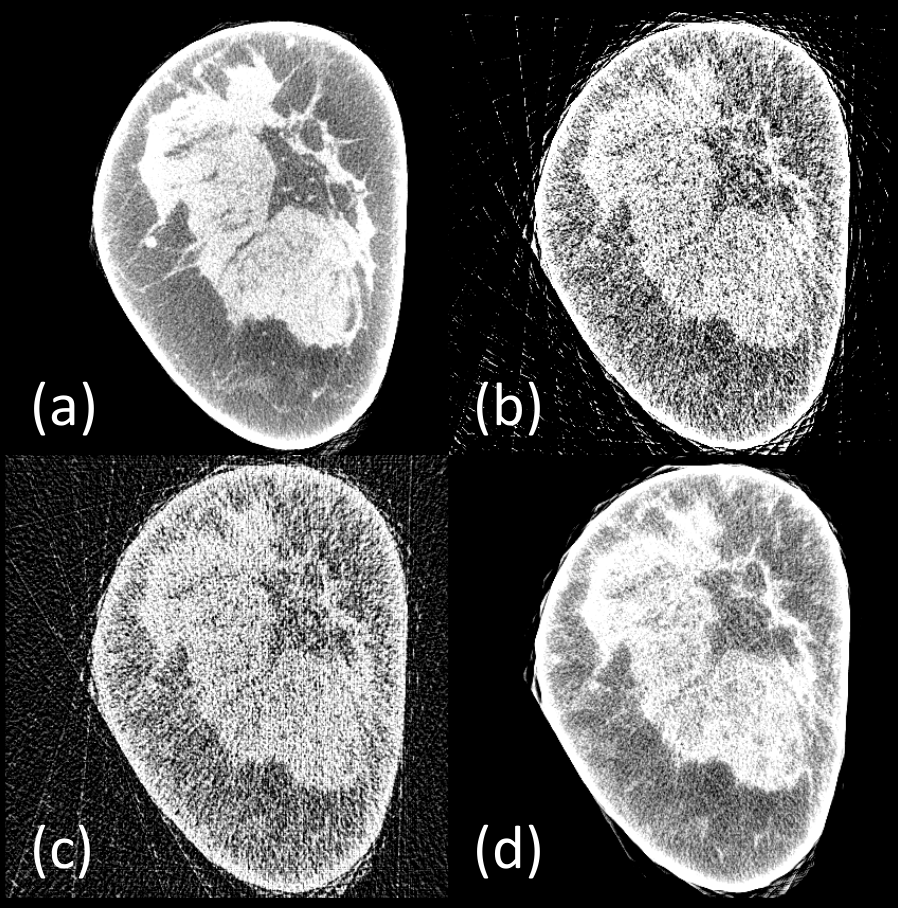}
\caption{Representative slice reconstructed using different methods for Koning dataset for directly testing on 30-view data using a trained 75-view network. (a) Ground-truth, (b) FDK from 30-view cone-beam data, (c) FBP from 200-view simulated parallel-beam, (d) DEER-Lite. The display window is [-200, 200] HU. Note that the final reconstructions were post-processed to remove irrelevant structures outside the field of view.}
\label{fig:ON_30_200_result}
\end{figure}

Since the trainable parameters in DEER-Lite are view-independent, it can be applied to other numbers of views directly without further re-training. To show that DEER-Lite can be used for reconstructions in different few-view conditions, the network (which was trained on 75-view data) was tested for 30-view reconstructions. The testing details are the same as described above. First, few-view reconstructed volumes were produced from 30-view cone-beam projections using the FDK algorithm. Then, 200-view parallel-beam projections were estimated from the reconstructed volumes (we simulated 200-view parallel-beam projections instead of 150-view to show that DEER-Lite can handle other numbers of views directly). Note that because the reconstructed image from the back-projection part in DEER (and its variances) is a summation of back-projected profiles at different view angles, the magnitude of the learned reconstructed image will be changed with different numbers of views. Therefore, simple linear re-scaling is necessary to adjust the magnitude of the reconstructed image from the back-projection part according to the view number. Lastly, the trained DEER-Lite model, which was trained for 75-view reconstructions, was directly used to produce 30-view results. A representative slice is shown in Fig. \ref{fig:ON_30_200_result}. The reconstructed image is compared with the ground-truth image, the FDK result from 30-view cone-beam data, and the FBP result from simulated 200-view parallel-beam data. The corresponding quantitative measurements are shown in Table \ref{table:ON_30_quan}.

\begin{table}[!h]
\centering
\caption{Quantitative assessment on DEER-Lite ($\mathrm{MEAN}\pm \mathrm{STD}$) for directly testing on 30-view data using a trained 75-view network. The measurements were obtained by averaging the values on the testing dataset.}
\resizebox{0.45\textwidth}{!}{
\begin{tabular}{ccc}
\toprule
     &  FDK     &      DEER-Lite         \\
\midrule
PSNR & $22.1137\pm6.3728$ & $28.0983\pm8.4834$ \\
SSIM & { }{ }$0.7795\pm 0.1485$ & { }{ }$0.9018\pm0.0985$   \\
MAE  & { }{ }$0.0289\pm0.0270$ & { }{ }$0.0148\pm0.0167$   \\
\bottomrule
\end{tabular}
}
\label{table:ON_30_quan}
\end{table}

\section{Discussions} \label{sec::discussion}

Breast CT improves the detection and characterization of breast cancer, with the potential to become a primary breast screening tool. Our Deep Efficient End-to-end Reconstruction (DEER) network has been shown to be feasible, with a great potential for further improvements for few-view CT image reconstructions. This is achieved by directly mapping sinogram data to CT images, lowering the involved x-ray radiation dose well under the FDA threshold in a data-driven fashion. Also, the DEER network improves the computational complexity by orders of magnitude relative to the state of the art networks that map raw tomographic data to reconstructed images. 

This paper has introduced a novel approach for reconstructing CT images directly from under-sampled projection data, referred to as DEER. This approach is featured by (1) an adjustable network-based reconstruction algorithm, (2) the Wasserstein GAN framework for optimizing parameters, (3) a convolutional encoder-decoder network, and (4) an experimentally optimized objective function. The proposed end-to-end strategy has been applied to learn the mapping from sinogram domain to image domain, requesting significantly less computational burden than prior arts. Zhu \etal \cite{zhu_image_2018} published the first method for learning a network-based reconstruction algorithm for medical imaging. They used several fully-connected layers to learn the mapping between MRI raw-data and an underlying image directly. But their method poses a severe challenge for reconstructing normal-size images due to extremely large matrix multiplications in the fully-connected layers. Additionally, even though this technique could be applied to CT images, using fully-connected layers to learn the mapping does not use the full information embedded in the sinograms, resulting in redundant network parameters. iCT-Net \cite{li_learning_2019} utilizes angular information and reduces the number of parameters from $\mathcal{O}(N^4)$ to $\mathcal{O}(N^2\times N_d)$. Instead of feeding the whole sinogram directly into the network, iCT-Net uses $N_d$ fully-connected layers, each takes a single projection and reconstructs a corresponding intermediate image component. DEER takes full advantage of all the information embedded in the sinograms by utilizing the angular information similar to what the iCT-Net does and also assuming every single point in the sinogram is only related to reconstructing pixels along the associated X-ray path. The proposed method is a fully learnable mapping from the sinogram to the image domain. By taking advantage of the geometric constraint imposed by line integrals, our network is much more efficient than other end-to-end methods. The intuition of DEER is to learn a better filtering and back-projection process using deep-learning approaches. With this intuition and the novel network-based end-to-end method, DEER can be trained with $\mathcal{O}(N^2\times N_v)$ parameters. Moreover, with this design, the reconstruction process can be learned with as few as $\mathcal{O}(N)$ parameters if all the angles and ray-tracing lines share the same training parameters during the back-projection procedure. If the network only uses $\mathcal{O}(N)$ parameters (which means parameters are not view-dependent), the proposed back-projection mechanism could be applied to other numbers of views directly even without re-training due to the learned ray-tracing idea.

The method proposed in this paper can be easily extended to the clinical cone-beam setting by inputting 3D cone-beam data to the network and trying to learn the mapping between 3D cone-beam data and 3D image volumes directly in a similar manner as long as the GPU memory is sufficiently large. Since few-view artifacts, calcified deposits, and subtle details appear in 3D geometry, 3D networks can capture full spatial context and has a great potential to improve image quality \cite{xie_deep_2019, xie_3d_2020}. Nevertheless, training 3D networks directly from cone-beam data is not feasible on our current hardware. Once we have a workstation with larger GPU memory, by addressing the divergent cone-beam geometry and ray-specific varying weights, our DEER network can be modified and re-trained to handle cone-beam projection data directly for cone-beam image volume reconstruction.

Moreover, the proposed method can be easily implemented into other imaging modalities, such as single-photon emission computed tomography (SPECT) and positron emission tomography (PET) since their forward imaging models are similar to CT and their memory requirements are much more manageable. Translating the DEER network into SPECT and PET could have a particularly high impact in applications of organ-specific scanners with special geometry and new scanner designs using the same number of detector modules to cover a larger FOV.

In conclusion, we have presented a novel network-based reconstruction algorithm for few-view CT reconstruction. The proposed method outperforms existing deep-learning-based methods with significantly less memory burden and higher computational efficiency. In the future, we plan to further improve this network for direct cone-beam CT 3D reconstruction and other imaging modalities, and translate it into clinical applications.

\bibliographystyle{ieeetr}
\bibliography{main.bib}

\end{document}